\documentclass[aps,pra,twocolumn,superscriptaddress,amsmath,reprint]{revtex4-1}

\usepackage{graphicx}
\usepackage{amssymb}
\usepackage{gensymb}
\usepackage{footmisc}
\usepackage{booktabs}
\usepackage{multirow}
\usepackage[colorlinks,urlcolor=blue,citecolor=blue,linkcolor=blue]{hyperref}
\usepackage{color}

\begin{document}
\title{Synthetic Hall ladder with tunable magnetic flux}
\author{Jeong Ho Han}
\email{jeongho.han@kriss.re.kr}
\affiliation{
Korea Research Institute of Standards and Science, Daejeon 34113, Korea
}
\author{Dalmin Bae}
\affiliation{
Department of Physics and Astronomy, Seoul National University, Seoul 08826, Korea
}
\affiliation{
Center for Correlated Electron Systems, Institute for Basic Science, Seoul 08826, Korea }

\author{Yong-il Shin}
\email{yishin@snu.ac.kr}
\affiliation{
Department of Physics and Astronomy, Seoul National University, Seoul 08826, Korea
}
\affiliation{
Center for Correlated Electron Systems, Institute for Basic Science, Seoul 08826, Korea }
\affiliation{
Institute of Applied Physics, Seoul National University, Seoul 08826, Korea
}

\date{\today}

\begin{abstract}
We describe a synthetic three-leg Hall ladder system with a tunable magnetic flux for neutral $^{173}$Yb atoms in a one-dimensional optical lattice. The ladder legs are formed by three hyperfine ground spin states of the atoms, and the complex interleg links are generated through Raman couplings between the spin states using multiple laser beams. The effective magnetic flux through a ladder plaquette, $\phi$, is controlled by the angles of the Raman laser beams with the lattice axis. We investigate the quench dynamics of the Hall ladder system for $\phi\approx\frac{\pi}{3}, \frac{\pi}{2},$ and $\frac{2\pi}{3}$ after a sudden application of the Raman coupling in various interleg link configurations. The semi-classical trajectory of the atoms in the plane of the spin composition and lattice position exhibits the characteristic motion for the effective magnetic field. In a tube configuration with the three legs cyclically linked, the quench evolution was observed to be substantially damped, which is attributed to the random flux threading the Hall tube.

\end{abstract}

\pacs{XX.XX.XX}
\maketitle

\section{Introduction}

Understanding topological states of matter is among the major challenges of condensed matter physics, with relevance for new classes of materials beyond the conventional symmetry-breaking paradigm~\cite{Hasan2010}. Owing to the capabilities to engineer tunneling amplitudes and address individual lattice sites, ultracold atoms in optical lattices represent a promising platform for exploring such topological phases in a defect-free environment~\cite{Goldman2016,Cooper2019}. One of the prominent examples is the quantum simulation of a Harper-Hofstadter (HH) Hamiltonian, which is the paradigmatic model for quantum Hall physics. The HH Hamiltonian was first achieved in two-dimensional (2D) optical lattice systems using laser-assisted tunneling~\cite{Aidelsburger2013, Miyake2013} and further realized in novel forms based on so-called synthetic dimensional lattices~\cite{Celi2014,Mancini2015,Stuhl2015}. In the synthetic dimension framework, a set of atomic internal states is employed as virtual lattice sites in an extra dimension and an effective gauge potential is generated by engineering the couplings between the states. Various Hall ladders~\cite{Livi2016,An2017-1,Kolkowitz2017} and topological lattice systems~\cite{An2018,Meier2016,Kang2020,Kanungo2021} were successfully demonstrated. Further, this synthetic dimension approach enables Hall effects to be studied even under hypothetical situations such as in tube geometries~\cite{Han2019,Chalopin2020,Liang2021} and in four dimensions~\cite{Lohse2018}. 

In our recent work~\cite{Han2019}, we constructed a synthetic three-leg Hall ladder for fermionic $^{173}$Yb atoms in a one-dimensional (1D) optical lattice, where three hyperfine ground spin states of atoms were utilized as the indices of the three legs and coupled to each other through two-photon Raman transitions. The linking structure of the ladder was altered by controlling the number of activated Raman couplings, including a tube configuration where the three spin states were cyclically coupled. In this study, a Hall ladder with an effective magnetic flux per unit plaquette of $\phi=2\pi/3$ was investigated as a representative case, featuring the existence of a symmetry-protected 1D topological phase~\cite{Barbarino2018}. Such a Hall tube system is known to belong to a 1D quasicrystal, exhibiting a fractal energy spectrum such as Hofstadter's butterfly for varying $\phi$ and carrying a number of topologically nontrivial states. Therefore, a natural extension for the Hall ladder system is to attain magnetic flux tunability for a full study of the quantum Hall physics. 

In this paper, we present a detailed description of the synthetic three-leg Hall ladder system and experimentally demonstrate the flux tunability of the system. We investigate the quench evolution of the system after a sudden application of the Raman couplings for different ladder structures and effective magnetic fluxes. The semi-classical trajectories of the atoms are reconstructed in the synthetic space composed of real and synthetic dimensions, and the effect of the tunable magnetic flux is demonstrated by comparing the measured dynamics with the numerical simulation results calculated from the corresponding Bloch equations. For the periodic boundary ladder, we observe a large damping in the spin evolution. In the tube geometry, there is an additional magnetic flux, which pierces the tube cross-section and affects the quasi-momentum along the circumference of the tube~\cite{Laughlin1981,Luo2020,Liang2021}. We attribute the large damping to the random flux through the tube owing to our Raman beam configuration, which is also compared with numerical simulation results. We discuss various options for controlling the longitudinal flux in the experiment.

The remainder of this paper is structured as follows. In Sec.~\ref{sec:model}, we introduce a model Hamiltonian governing the Hall ladder system for open and periodic boundary conditions. In Sec.~\ref{sec:exp}, we describe the experimental protocol to generate the synthetic Hall ladder system using neutral atoms in a 1D optical lattice. In Sec.~\ref{sec:result}, we present the quench experiments and discuss their results. Finally, some concluding remarks and outlooks are provided in Sec.~\ref{sec:conclusion}.

\section{Theoretical model}\label{sec:model}

\begin{figure}
\includegraphics[width=8.4cm]{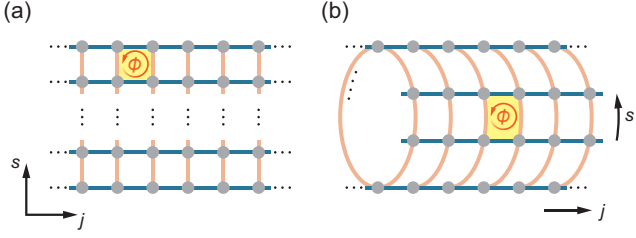}
\caption{Hall ladders with a uniform magnetic flux $\phi$ per plaquette. The ladder consists of a finite number of legs, indexed by $s$ with (a) open boundaries on its lateral sides or (b) a periodic boundary to form a cylindrical geometry. $j$ labels the lattice site along the leg direction.\label{fig:ladder}}
\end{figure}

A Hall ladder is a 2D square lattice system in the presence of a gauge potential, having a finite width for one dimension. We consider non-interacting spinless fermions in a Hall ladder, whose sites are indexed by $(j,s)$ with $j\in \mathbb{Z}$ and $s\in\{1,\cdots,\rho\}$ (Fig.~\ref{fig:ladder}). In the tight-binding approximation, the system's Hamiltonian is given by
\begin{equation} \label{eq:tb1}
\begin{aligned}
\hat{H}/\hbar&=\sum_{j,s} \left[-t_x  \hat{a}^\dagger_{j+1,s}\hat{a}_{j,s} + \frac{\Omega_{s}}{2} e^{i\phi j}\hat{a}^\dagger_{j,s+1}\hat{a}_{j,s} + \mathrm{h.c.} \right],
\end{aligned}
\end{equation}
where $\hat{a}_{j,s}$ ($\hat{a}^\dagger_{j,s}$) is the annihilation (creation) operator for the localized state on site $(j,s)$. In addition, $t_x$ and $\Omega_{s}$ are tunneling amplitudes between adjacent lattice sites, where we assume a real $t_x$ and complex-valued $\Omega_{s}$ in general. The tunneling along the $s$ direction has a $j$-dependent Peierls phase, resulting in a net phase $\phi$ for a loop around a unit plaquette, which corresponds to a magnetic flux per plaquette, $\Phi_B=(\phi/2\pi) \Phi_0$ with $\Phi_0$ being the magnetic flux quantum. Taking the transformation of $\hat{b}^\dagger_{q,s}=L^{-1/2}\sum_{j} \hat{a}^\dagger_{j,s} e^{i[q+(s-1)\phi]j}$ (where $L$ is the length of the ladder), $\hat{H}$ is expressed as
\begin{equation}\label{eq:tb2}
\begin{aligned}
\hat{H}/\hbar=\sum_{q,s} \big[ & h_q(\phi,s) \hat{b}^{\dagger}_{q,s}\hat{b}_{q,s} \\
&+\frac{\Omega_{s}}{2} \hat{b}^{\dagger}_{q,s+1}\hat{b}_{q,s} + \frac{\bar{\Omega}_{s-1}}{2}  \hat{b}^{\dagger}_{q,s-1}\hat{b}_{q,s}\big],\\
\end{aligned}
\end{equation}
where $q$ is the quasimomentum along $j$, and $h_q(\phi,s)=-2t_x\cos [q+(s-1)\phi]$ represents the energy dispersion of leg $s$ with momentum shifted by $(s-1)\phi$. The off-diagonal terms infer that the Hall ladder can be viewed as a coupled $\rho$-leg system.

\begin{figure}
\includegraphics[width=8.5cm]{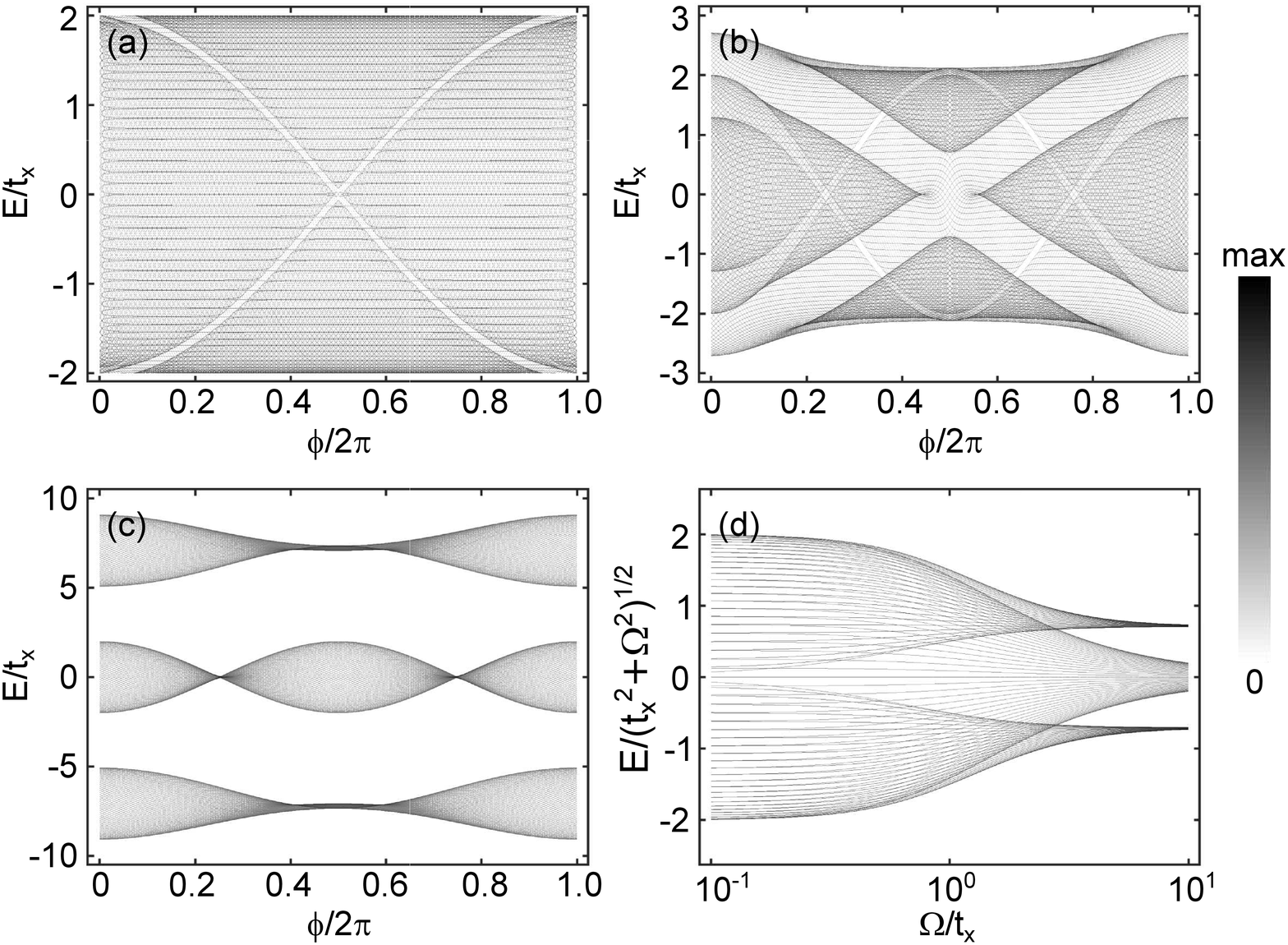}
\caption{Density of states (DoS) of open 3-leg Hall ladders as functions of magnetic flux $\phi$ for (a) $\Omega/t_x=0.1$, (b) 1, and (c) 10. (d) Evolution of DoS as a function of $\Omega/t_x$ for $\phi=\pi$.\label{fig:dos}}
\end{figure}

In the case of $\Omega_{\rho}=0$, the Hall ladder has an open boundary with edges at $s=1$ and $\rho$ [Fig.~\ref{fig:ladder}(a)]. The Bloch Hamiltonian is given in a $\rho\times \rho$ matrix form by
\begin{equation} \label{eq:tb3}
\begin{aligned}
&\hat{H}_{q}/\hbar\\
&=\begin{pmatrix}
h_{q}(\phi,1) & \frac{\Omega_{1}}{2} & 0 & \cdots & 0 \\ 
\frac{\bar{\Omega}_{1}}{2}  & h_{q}(\phi,2) & \frac{\Omega_{2}}{2}  & \cdots & 0\\ 
0 & \frac{\bar{\Omega}_{2}}{2}  & \ddots &  & \vdots\\ 
\vdots & \vdots &  & \ddots & \frac{\Omega_{\rho-1}}{2} \\ 
0 & 0 & \cdots & \frac{\bar{\Omega}_{\rho-1}}{2}  & h_{q}(\phi,\rho)
\end{pmatrix}.
\end{aligned}
\end{equation}
In Fig.~\ref{fig:dos}, we display the density of states of an equilateral ladder, i.e., $\Omega_{s}=\Omega$, as a function of $\phi$ for various $\Omega/t_x$. As the interleg coupling strength increases, energy gaps are opened, indicating strong avoided crossings. All bands in the open boundary Hall ladder have a vanishing 1D winding number. The topological bands can be formed when additional hopping terms are introduced, such as next nearest neighbor tunneling, i.e., diagonal links between the legs~\cite{Hugel2014}.

When $\Omega_{\rho}\neq0$, the system has a periodic boundary, forming a tube geometry [Fig.~\ref{fig:ladder}(b)]. The periodic condition is achieved only if $\rho\phi/(2\pi)=P/Q$, where $P$ and $Q$ are co-prime integers. This requirement corresponds to the wavefunction of a particle being single-valued after $Q$ loops around the tube. With $\phi/(2\pi) = P'/Q'$ (where $P'$ and $Q'$ are co-prime integers), we have $\rho P'/Q'=P/Q$ and $Q=\mathcal{D}/\rho$ with $\mathcal{D}=\mathrm{lcm}(\rho,Q')$. The Bloch Hamiltonian of the periodic system is given by
\begin{equation} \label{eq:tb4}
\begin{aligned}
&\hat{H}_{q}/\hbar\\
&=\begin{pmatrix}
h_{q}(\phi,1) & \frac{\Omega_{1}}{2} & 0 & \cdots & \frac{\bar{\Omega}_{\mathcal{D}}}{2} \\ 
\frac{\bar{\Omega}_{1}}{2}  & h_{q}(\phi,2) & \frac{\Omega_{2}}{2}  & \cdots & 0\\ 
0 & \frac{\bar{\Omega}_{2}}{2}  & \ddots &  & \vdots\\ 
\vdots & \vdots &  & \ddots & \frac{\Omega_{\mathcal{D}-1}}{2} \\ 
\frac{\Omega_{\mathcal{D}}}{2}  & 0 & \cdots & \frac{\bar{\Omega}_{\mathcal{D}-1}}{2}  & h_{q}(\phi,\mathcal{D})
\end{pmatrix},
\end{aligned}
\end{equation}
where periodic labels along $s$ are adopted, i.e., $h_q(\phi,s+\rho)=h_q(\phi,s)$ and $\Omega_{s+\rho}=\Omega_{s}$. In general, the system exhibits a $\mathcal{D}$-band energy spectrum.

In a tube geometry, we need to consider an additional magnetic flux that threads the longitudinal direction of the tube, which is defined as $\phi_\mathrm{L} = \mathrm{arg}(\Omega_{1}\cdots\Omega_{\rho})$. For the equilateral case with $\Omega_{s}=\Omega=|\Omega|e^{i \phi_\mathrm{L}/\rho}$, the physical meaning of $\phi_\mathrm{L}$ is clear. The translational symmetry along the tube circumference defines the quasi-momentum $p=0, \frac{2\pi}{\rho}, \cdots, \frac{2\pi}{\rho}(\rho-1)$ and the wavefunction fulfills the Bloch theorem. Following a substitution $\hat{b}_{q,s} = e^{ips} \hat{c}_{q,s}$, $\hat{H}$ is expressed as
\begin{equation} \label{eq:tb5}
\begin{aligned}
\hat{H}/\hbar=\sum_{q,s} \big[ & h_q(\phi,s) \hat{c}^{\dagger}_{q,s}\hat{c}_{q,s} \\
&+\frac{\Omega}{2} e^{-ip} \hat{c}^{\dagger}_{q,s+1}\hat{c}_{q,s} + \frac{\bar{\Omega}}{2} e^{ip} \hat{c}^{\dagger}_{q,s-1}\hat{c}_{q,s}\big],\\
\end{aligned}
\end{equation}
where the form of the off-diagonal terms shows that the effect of the longitudinal flux $\phi_\mathrm{L}$ is a shift of the momentum $p$ to $p'=p-(\phi_\mathrm{L}/\rho)$, analogous to the Aharonov-Bohm effect for a charged particle. According to Laughlin's argument, an adiabatic change in $\phi_\mathrm{L}$ results in topological charge pumping along the leg direction in the system, as demonstrated in recent experiments~\cite{Fabre2021}. Note that the total magnetic flux piercing the cross section of the tube at $j$ is $\Phi_j=\rho \phi j + \phi_\mathrm{L}$.

\begin{figure}
\includegraphics[width=7cm]{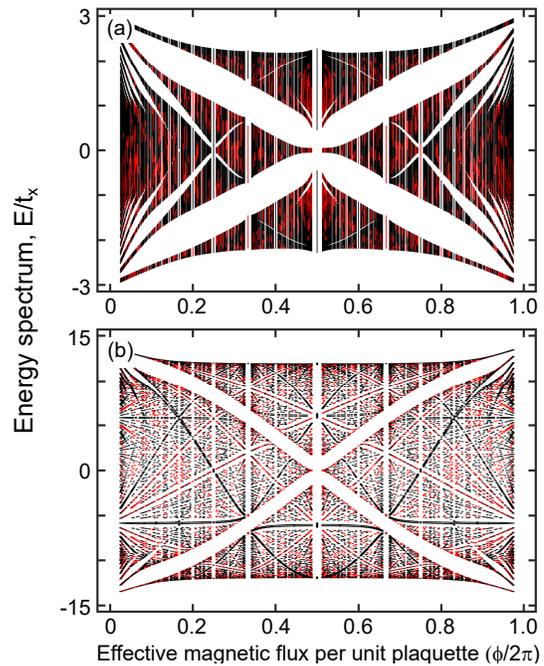}
\caption{Hofstadter's butterfly emergent in periodic 3-leg Hall ladder. Energy spectra of the system for (a) the isotropic tunneling strength $\Omega/t_x = 1$ and (b) an anisotropic case $\Omega/t_x = 11.7$. The topologically nontrivial energy bands with nonzero Zak phase are indicated by red lines.\label{fig:1dbutterfly}}
\end{figure}

In Fig.~\ref{fig:1dbutterfly}, we display the energy spectra of the equilateral Hall ladder with $\rho=3$ as a function of $\phi$ for various $\Omega/t_x$ and $\phi_\mathrm{L}=0$. Because Eq.~(\ref{eq:tb4}) satisfies the recursive Harper equation, the spectra exhibit a self-similarity like Hofstadter's butterfly, which is a hallmark of quasi-crystals~\cite{Rajagopal2019, Dean2013}. Figure~\ref{fig:1dbutterfly}(b) shows an anisotropic case for $\Omega/t_x =11.7$, which is close to the value for the experiments that follow. In Ref.~\cite{Barbarino2018}, it was shown that $\hat{H}_{q}$ for $\phi=2\pi/\rho$ ($\rho$ is odd) and $\phi_\mathrm{L}=0$ embeds a 1D topologically insulating phase with nontrivial winding number protected by generalized inversion symmetry. From Zak phase calculations, we found that there are more topologically nontrivial bands for different $\phi$, which are indicated as red lines in Fig.~\ref{fig:1dbutterfly}. When the system is project onto the Brillouin zone spanned by $q$ and $p$, it is also characterized by a 2D topological invariant, i.e., a Chern number~\cite{Abanin2013}.

\section{Experimental setup}\label{sec:exp}

\subsection{Sample preparation}

\begin{figure}
\includegraphics[width=7.5cm]{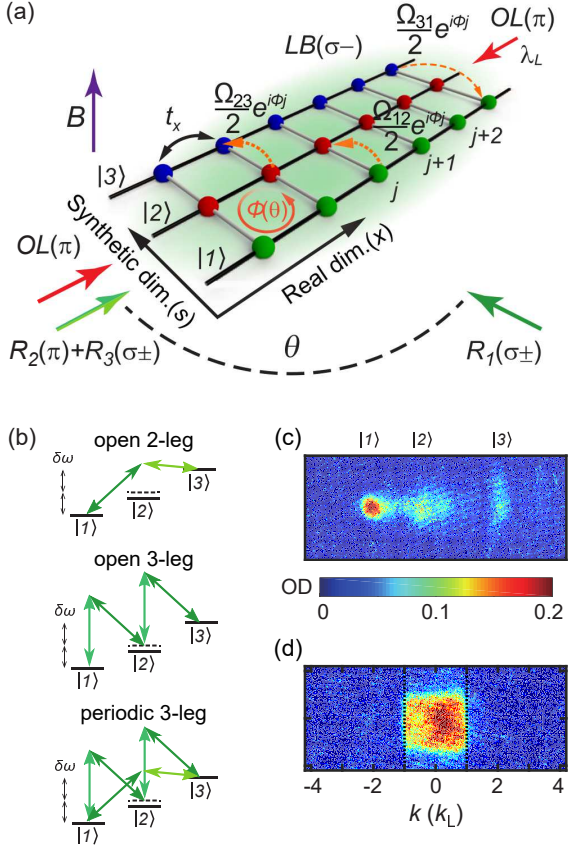}
\caption{(a) Sketch of the experimental setup. Ultracold fermionic atoms of $^{173}$Yb are trapped in 1D optical lattice. A set of Raman beams drive the complex interleg tunnelings between the synthetic lattice sites, generating the effective magnetic flux $\phi(\theta)$ perpendicular to the $xs$-plane. (b) The linking structure of the synthetic Hall ladder is determined by the Raman coupling schemes in the three lowest atomic spin states. (c) Exemplary images of the atoms obtained after optical Stern-Gerlach detection and (d) adiabatic band mapping, where the samples are quenched for (c) 0.1 and (d) 0.15~ms, respectively. \label{fig:scheme}}
\end{figure}

We begin by preparing a degenerate Fermi gas of $^{173}$Yb spin-polarized in a ground state $|F,m_F\rangle=|5/2,-5/2\rangle$ confined in a three-dimensional optical lattice potential, as described in our previous experiment~\cite{Han2019}. A typical sample contains $\approx 1.2\times10^5$ atoms with a residual spin fraction ($m_F \neq -5/2$) of less than 3\% and has a temperature $\approx 0.35 T_F$, where $T_F$ is the Fermi temperature of the trapped sample. The optical lattice potential is formed by three pairs of interfering laser beams at a wavelength $\lambda_L = 532$~nm, resulting in lattice constants $d_{x,z} = \lambda_L/2$ and $d_y = \lambda_L/\sqrt 3$. The final lattice depths are $(V_x,V_y,V_z)=(4.5,20,20)E_{L,\alpha}$, where $E_{L,\alpha}=h^2/8md^2_\alpha$, $\alpha\in\{x,y,z\}$ are the corresponding recoil energies, $h$ is the Planck constant, and $m$ is the atomic mass. Within our experiment timescale (a few milliseconds), the high lattice depths $V_{y}$ and $V_z$ strongly suppress the atomic motion along $y$ and $z$ between the adjacent individual planes, generating effective 1D lattice with a tunneling rate of $t_x=2\pi\times298$~Hz. The trapping frequencies of the overall harmonic confinement are estimated to be $(\omega_x, \omega_y, \omega_z) = 2\pi\times(57, 40, 130)$~Hz at the end of the preparation sequence.

\subsection{Realization of synthetic Hall ladders}

We realize a 2D Hall ladder by hybridizing the 1D optical lattice with synthetic tunneling between the internal atomic spin states. The sites of the synthetic dimension are provided by a subset of $^1$S$_0$ ground spin states, which we denote as $|1\rangle\equiv|m_F=-5/2\rangle$, $|2\rangle\equiv|m_F=-3/2\rangle$, and $|3\rangle\equiv|m_F=-1/2\rangle$, constituting the legs of the ladder system. The ladder rungs are generated by two-photon Raman transitions between these spin states. As shown in Figure~\ref{fig:scheme}(a), three Raman beams $R_{1,2,3}$ propagating in the $xy$ plane are irradiated on the sample with wavevectors $\vec{k}_{r1}=k_R(\cos{\theta}\hat{x}+\sin{\theta}\hat{y})$ and $\vec{k}_{r2}=\vec{k}_{r3}=k_R\hat{x}$, respectively. The polarization directions of the Raman beams are horizontal for $R_{1,3}$ (responsible for $\sigma^\pm$) and vertical for $R_2$ (responsible for $\pi$) toward the $xy$ plane. To lift the ground state degeneracy, an external magnetic field of $B=153~$G is applied along $z$. In this configuration, a Raman pair $R_2R_1$($\pi\sigma$) drives the couplings $|1\rangle$$\leftrightarrow$$|2\rangle$ and $|2\rangle$$\leftrightarrow$$|3\rangle$ changing the quantum number by $\Delta m_F = 1$, whereas $R_1R_3$($\sigma\sigma$) generates the coupling between $|1\rangle$$\leftrightarrow$$|3\rangle$ with $\Delta m_F = 2$.

In a synthetic dimensional frame, the Raman transition between $|s\rangle$ and $|s'\rangle$ can be interpreted as a interleg tunneling with complex amplitude $\Omega_{ss'}e^{i\phi j}/2$, where $\Omega_{ss'}$ is the Rabi frequency and $j$ is the site index for the real lattice. We adjust the Raman beam intensities to satisfy $\Omega_{12}=\Omega_{31}=11.7t_x$; we emphasize that $\Omega_{12}/\Omega_{23}$ equals almost unity (within 2\%) because both transitions originate from the same Raman beam pair, $R_2R_1$. The momentum transfer owing to the Raman photons imprints the spatial phase modulation on the synthetic tunneling amplitude,
\begin{equation}
\phi(\theta)=[(\vec{k}_{r2,r3}-\vec{k}_{r1})\cdot\hat{x}]d_x=k_R d_x (1-\cos{\theta}),
\end{equation}
which is experimentally tunable by varying $\theta$. The atoms hopping around the unit plaquette acquire a uniform phase $\phi$. In this manner, we generate a synthetic Hall ladder system governed by Eq.~(\ref{eq:tb1}). 

The essence of producing a ladder geometry is the precise manipulation of atomic energy levels $\nu_s$, because we employ an energy mismatch to isolate the ladder manifold from the residual spin states. In Ref.~\cite{Stuhl2015}, unwanted Raman transitions are suppressed by the second-order Zeeman shift in $\nu_s$; in Ref.~\cite{An2017-1,An2018,Meier2016}, the separation originates from the nonlinearity of the free particle energy spectrum. In this experiment, we exploit the state-dependent light shift by dressing an additional $\sigma^-$ polarized laser beam on the atoms to detach the superfluous link, particularly between $|3\rangle$ and $|4\rangle$. This laser beam has an intensity of 8.5 mW cm$^{-2}$ and is detuned $-70$~MHz from the $|^1S_0, F=5/2\rangle$$\rightarrow$$|^3P_1,F'=7/2\rangle$ resonance, adding a positive (negative) light shift for $\nu_1$, $\nu_2$, and $\nu_3$ ($\nu_4$, $\nu_5$, and $\nu_6$), respectively, which drives the two-photon transition between $|3\rangle$ and $|4\rangle$ off-resonant. Along with the light shifts from  $R_{1,2,3}$, the final experimental condition is $\left(\xi_{1},\xi_{2},\xi_{3},\xi_4,\xi_5\right)$$\approx$$\left(0,-0.28,0,-2.52, 2.25\right)\Omega_{12}$, where $\xi_s = \left(\nu_s - \nu_1\right)-(s-1)\delta \omega$ is the detuning of $|s\rangle$ from the two-photon resonance, and $\delta\omega=2\pi \times 30.4$~kHz. In this scheme, $R_{1,2,3}$ are dominated by $\sigma^\pm$ beams, which generate light shifts in atomic energy levels with a small asymmetry in $\xi_2$. This gives rise to a minor leakage channel between $|2\rangle$ and $|4\rangle$, causing a typical atom loss of less than 8$\%$ after a $1$-ms evolution.

\begin{center}
\begin{figure*}
\includegraphics[width=16.5cm]{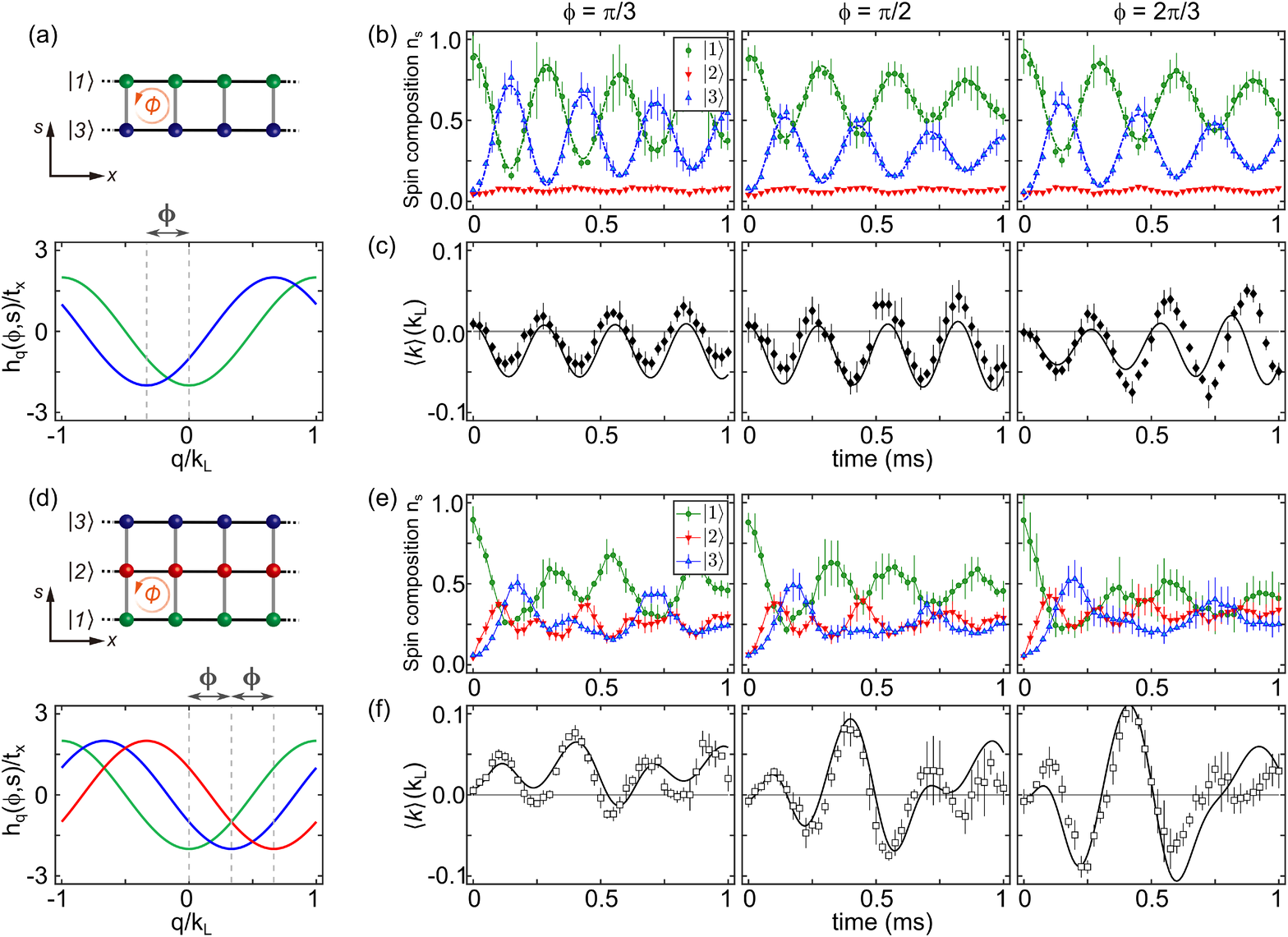}
\caption{Time dependence of the spin composition $n_s$ and total lattice momentum $\langle k \rangle$ of open boundary synthetic Hall ladders for different values of an effective magnetic flux. (a) Schematic, (b) $n_s$, (c) $\langle k \rangle$ of the 2-leg ladder, (d) schematic, (e) $n_s$, and (f) $\langle k \rangle$ of the 3-leg ladder are shown. The dashed lines are fitted curves using a damped oscillator model [Eq.~(\ref{eq:dampingfit})] and the solid lines indicate the results from the numerical simulation obtained by solving the Bloch equations. Each data point comprises five independent measurements of the same experiment and the error bar is their standard deviation.\label{fig:openleg}}
\end{figure*}
\end{center}

The structure of the Hall ladder system is determined by the connectivity between the legs of the synthetic lattices. As depicted in Figure~\ref{fig:scheme}(b), we vary the number of legs by controlling the number of resonant Raman transitions employed in the system. To create a periodic boundary, we place the laser frequencies of $R_{1,2,3}$ at $\omega_1 = \omega$, $\omega_2 = \omega+\delta\omega$, and $\omega_3 = \omega-2\delta\omega$, where $\omega$ is the laser frequency blue-detuned by $1.97$~GHz from the $|^1S_0, F=5/2\rangle$$\rightarrow$$|^3P_1,F'=7/2\rangle$ transition. The two(three)-leg ladder under an open boundary condition can be realized by shifting $\omega_2$($\omega_3$) by $2\pi\times400$($-400$)~kHz from the Raman resonance, deactivating the corresponding interleg links $R_2R_1$($R_1R_3$), respectively. For a large detuning, the associated interleg couplings are effectively switched off but the light shifts for the Hall ladder formation are nearly unaffected.

\section{Results and Discussions} \label{sec:result}

\subsection{Quench dynamics for open boundary ladders}

Quench dynamics refers to the evolution of a system after a sudden change in the system parameters and it has been used as an essential tool in the study of topological lattice systems~\cite{Eisert2015}. Examples include reconstructing the band structure of a spin-orbit coupled lattice through Fourier spectroscopy~\cite{VCuriel2017}, the observation of dynamical vortices in a momentum space~\cite{Flaschner2018}, spin relaxation after quenching across the symmetry-protected topological phases~\cite{Song2018}, and the determination of topological invariant based on a post-quench evolution~\cite{Sun2018}. The quench dynamics of a system can also be exploited to extract macroscopic topological properties~\cite{Unal2016}. 

We conducted a quench experiment of the synthetic Hall ladders for various effective magnetic fluxes. Our experimental sequence is directed as follows. After preparing the atoms in $|1\rangle$ in a 1D optical lattice, we suddenly allow the interleg tunneling by illuminating the Raman beams on the samples for a variable time. The time dependence of the spin composition is measured through optical Stern-Gerlach detection~\cite{OSG}. Figure~\ref{fig:scheme}(c) displays an image of a spin-separated sample containing the atoms transferred from $|1\rangle$ to $|2\rangle$ and $|3\rangle$. Because the optical Stern-Gerlach process destroys the momentum profile of the atoms, we separately measure the lattice momentum distribution $n(k)$ using adiabatic band mapping in the samples under the same conditions~[Fig.~\ref{fig:scheme}(d)]. In the band mapping, slowly switching off the trapping lattice potential transforms the lattice momentum state $|q\rangle$ into the corresponding free-space momentum states $|k_s\rangle$, where the momentum $k_s$ of the spin $|s\rangle$ is related to $q$ as $k_s d_x = [q+(s-1)\phi]$ modulo $2\pi$ and $-k_L < k_s \leq k_L$ with $k_L=\pi/d_x$. To investigate the effective magnetic flux dependence of the quench dynamics, we carried out the experiment for three different fluxes, $\phi=\pi/3,\pi/2,$ and $2\pi/3$ by setting the angle of incidence of $R_1$ to $\theta=50^\circ, 60^\circ,$ and $72^\circ$, respectively.

We first investigate the quench dynamics in an open 2-leg Hall ladder, which is realized by deactivating the $|1\rangle$-$|2\rangle$ and $|2\rangle$-$|3\rangle$ interleg couplings. In Fig.~\ref{fig:openleg}(b), we plot the time evolution of the spin compositions for $\Omega_{12}/t_x=11.7$. During the evolution, the ladder demonstrates two-level Rabi dynamics between $|1\rangle$ and $|3\rangle$, whereas the population of $|2\rangle$ remains at a low fraction. From the measured $n(k)$, we obtain the total average momentum $\langle k \rangle=\int^{k_\mathrm{L}}_{-k_\mathrm{L}} k n(k)dk /\int^{k_\mathrm{L}}_{-k_\mathrm{L}} n(k) dk$, the time evolution of which is shown for a different $\phi$ in Fig.~\ref{fig:openleg}(c). We observe that $\langle k \rangle$ oscillates in the negative direction and its amplitude gradually increases over time. This originates from the interplay between the coupling structure and the initial momentum distribution. The $|k,1\rangle$ state is coupled to the $|k-\phi/d_x, 3\rangle$ for $k\in[-k_L+\phi/d_x,k_L]$, which drives the negative momentum recoil, whereas the $|k,1\rangle$ state is coupled to $|k+2 k_L-\phi/d_x,3\rangle$ for $k\in[-k_L,-k_L+\phi/d_x)$, which gives rise to a positive shift in lattice momentum owing to the lattice periodicity~[Fig.~\ref{fig:openleg}(a)]. Under our initial condition, the $s=1$ band is not completely filled and is more populated at near $k=0$, causing $\langle k \rangle$ to start oscillating in the negative direction. The gradual increase in the oscillation amplitude arises, because the average oscillation period for $k$ giving a positive shift differs from that for the other momentum range owing to different coupling strengths. For a higher $\phi$, the momentum range giving a positive shift in $\langle k\rangle$ is larger, and thus the growth in amplitude increases in speed~[Fig.~\ref{fig:openleg}(c)].

\begin{center}
\begin{figure*}
\includegraphics[width=15.0cm]{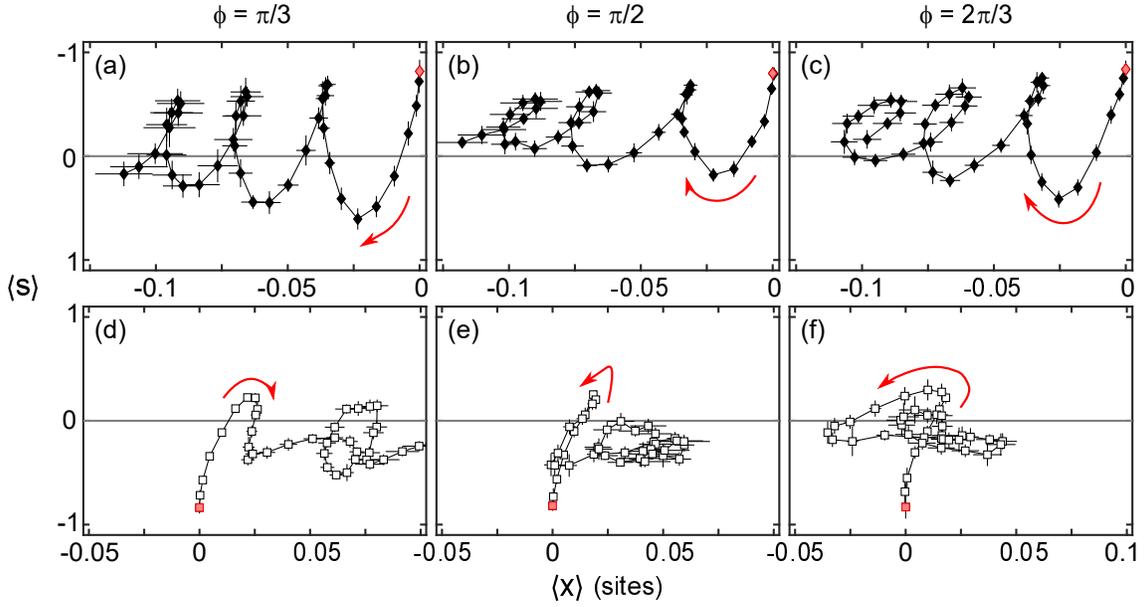}
\caption{Semi-classical trajectories of the atoms in the plane of the spin and real lattice positions for the quench dynamics of (a)-(c) the open boundary 2-leg and (d)-(f) 3-leg ladder systems for different values of effective magnetic flux. The initial positions are denoted by the red marks. \label{fig:trajectory}}
\end{figure*}
\end{center}

To validate our results, we perform a numerical simulation of the time evolution of the system by solving the Bloch equation,
\begin{equation} \label{eq:Blocheq}
i\hbar\frac{\partial}{\partial t} \Psi (q,t) = \hat{H}_q \Psi (q,t),
\end{equation}
where $\Psi(q,t)=[c_1,c_2,c_3]^{T}$ is the wavefunction of the fermions with probability amplitude $c_s(q,t)$ for spin $|s\rangle$. The atomic density for spin $s$ and momentum $k_s$ is given by $n_s(k_s,t)=|c_s(q,t)|^2$, and $n(k,t)=\sum_s n_s(k_s,t)$. We use the initial condition $\Psi(q,0) =[\sqrt{n_1(k_1,0)},0,0]$, where $n_1(k_1,0)$ is obtained by averaging the experimentally measured lattice momentum distributions of the spin-polarized atoms at $t=0$. In Fig.~\ref{fig:openleg}(c), the simulation results are displayed as solid lines, which correctly anticipate the time evolution in $\langle k \rangle$ for a different $\phi$. In the simulation, we also observe that the increasing slope of the oscillation envelope is reversed after $t>4$~ms.

Next, we investigate the quench dynamics in an open 3-leg Hall ladder, which is realized by deactivating the $|1\rangle$-$|3\rangle$ interleg coupling. The time evolutions of $n_s$ and $\langle k \rangle$ for the open 3-leg Hall ladder are displayed in Fig.~\ref{fig:openleg}(e) and \ref{fig:openleg}(f), respectively. In comparison with the 2-leg case, a $\phi$-dependence of the $\langle k \rangle$ oscillations dramatically appears with a large change in amplitude. For $\phi=\pi/3$, the dynamics of $\langle k \rangle$ evolves mostly on $\langle k \rangle >0$, whereas a change in sign of $\langle k\rangle$ occurs for $\phi=\pi/2$ and $2\pi/3$~[Fig.~\ref{fig:openleg}(f)]. This behavior originates from the multiple Raman processes involved in the 3-leg case and the fact that the large magnetic flux causes the atoms to be reflected at the Brillouin zone boundary by the Bloch oscillation~\cite{Han2019}. For example, the atoms initially located at $\langle k \rangle = 0$ experience two consecutive Raman transitions, which exerts a momentum recoil exceeding $k_L$ if $\phi \geq \pi/2$, forcing the atoms to shift forward $\langle k \rangle <0$ states at near $t\approx0.25$~ms. We found that numerical calculations including a phenomenological damping effect~\cite{damping} are in good agreement with the experiment results.

\subsection{Semi-classical trajectories for open boundary ladders}

In the presence of a magnetic field, a charged particle shows a characteristic cyclotron motion owing to the Lorentz force perpendicular to the velocity of a particle. For the particle in a Hall ladder system, a similar but more complicated motion is expected, because of the reflection from the lattice potential as well as the lateral edges. From the measurements, we reconstructed the trajectory of the average position of the atoms in the $xs$ plane, following the analysis described in Ref.~\cite{Mancini2015}. In brief, the atom in $|k\rangle$ travels with the group velocity $v_k={\partial\epsilon (k)}/{\hbar\partial k}$, where $\epsilon(k) = 2t_x [1-\cos(kd_x)]$ is the lattice dispersion of the ground band. We neglected the effect of the trapping potential, whose timescale is an order of magnitude slower than the experimental timescale $\approx 1$~ms. The average velocity of the atoms, $\langle v \rangle$, is obtained based on the density-weighted sum of the group velocity and the average position of the atoms, $\langle x \rangle$, is calculated by integrating the average velocity over time, i.e.,
\begin{equation}
\langle x(t) \rangle = \int_0^t \langle v(t) \rangle d\tau = \int_0^t \int_{-k_L}^{k_L} n(k,\tau) v_k ~dk d\tau,
\end{equation}
whereas the average position in $s$ is determined as $\langle s(t) \rangle=n_3(t)-n_1(t)$.

In Fig.~\ref{fig:trajectory}, we plot the reconstructed quench trajectory of the system in the $xs$ plane. The 2-leg ladder case [Fig.~\ref{fig:trajectory}(a)-\ref{fig:trajectory}(c)] shows a damped cyclotron motion truncated by the ladder edge. Noticeably, as $\phi$ increases, the cyclotron motion tilts toward the negative $x$-direction with growing overlapping regions. This is compatible with the observations in Figs.~\ref{fig:openleg}(c) and $\ref{fig:openleg}$(d), where the time dependence of momentum shows a rather increasing envelope, whereas the amplitude of the spin composition oscillations is slowly damped. 

The 3-leg ladder case reveals more intriguing trajectories [Fig.~\ref{fig:trajectory}(d)-\ref{fig:trajectory}(f)]. In the $\phi=\pi/3$ ladder, the atoms are shifted toward positive $x$-direction, because their $\langle k\rangle$ does not change its sign. In the $\phi=\pi/2$ and $2\pi/3$ cases, the atoms display bouncing motions, because of the Bloch oscillations during the cyclotron orbits. We emphasize that the trajectory is almost retraced back for $\phi=\pi/2$ at the first bounce, indicating the symmetric reflection at $k=k_L$ (corresponding to $2\phi=\pi$).

\begin{center}
\begin{figure}
\includegraphics[width=8.4cm]{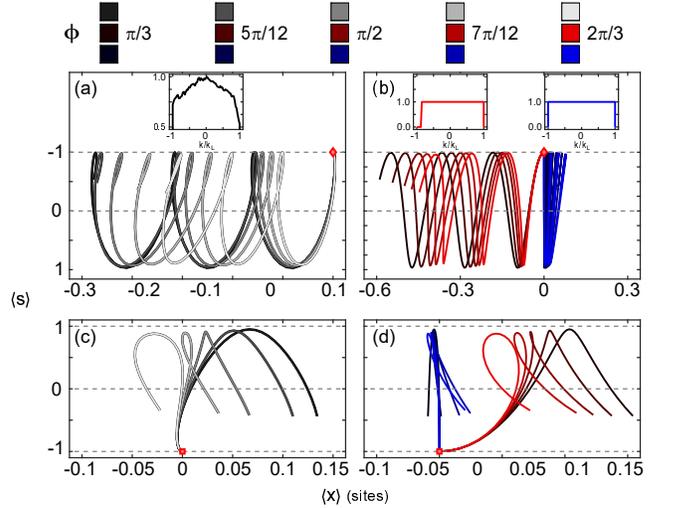}
\caption{Numerical simulation of the atomic trajectories for various magnetic flux after the quench at $\Omega_{31}/t_x=11.7$. The trajectories for (a) 2-leg ladder with measured initial momentum profile (black), (b) with partially occupied band (red) and completely filled band (blue), (c) 3-leg ladder with measured initial momentum profile (black), (d) with partially occupied band (red) and completely filled band (blue) are shown, respectively. Insets display the corresponding initial momentum profiles. \label{fig:sim_traj}}
\end{figure}
\end{center}

To further understand the observed behaviors, we numerically simulate the semi-classical trajectories of the atoms in the $xs$ plane and study their initial condition dependence (Fig.~\ref{fig:sim_traj}). We examined four different lattice momentum distributions, including $n_1(k_1,0)$ measured from the experiments, the case of a completely filled band, and the hypothetical case of a partially filled band with a complete depletion in $[-k_L,-0.85k_L]$.

The simulation for the 2-leg ladders demonstrates that under all three initial conditions the trajectories of the atoms gradually tilt toward the negative $x$-direction as $\phi$ increases, which is consistent with the experiment results. In Fig.~\ref{fig:sim_traj}(a), the case using the measured $n_1(k_1,0)$ shows a cyclotron motion similar to the trajectories from the experiment data, whereas in Fig.~\ref{fig:sim_traj}(b), the case of the completely filled band has $\langle v(0)\rangle=0$ and exhibits a trajectory confined along $\langle s \rangle$ at the initial times. For the case of the partially filled band, the trajectories proceed toward the negative $x$-direction from the beginning, because the depleted population leads to finite $\langle v(0)\rangle\neq0$.

The simulation for the 3-leg ladders also agrees with the experiment results in terms of the $\phi$ dependence in the winding directions of the trajectories in the $xs$ plane. As $\phi$ increases, the atomic trajectory bends counter-clockwise, regardless of the initial momentum distributions [Fig.~\ref{fig:sim_traj}(c)]. Similar to the 2-leg case, the completely filled band shows a motion localized within a small region owing to $\langle v(0)\rangle=0$, whereas the partially filled bands with an asymmetric occupation leads to spiral trajectories [Fig.~\ref{fig:sim_traj}(d)]. In both the simulation and experiment, the semi-classical trajectories for open boundary Hall ladders demonstrate strong spin-momentum locking, which implies that the atoms experience a force perpendicular to their velocity, similar to the Lorentz force acting on a charged particle in a magnetic field.

\subsection{Characterization of damping effect in periodic boundary ladders}

\begin{center}
\begin{figure}
\includegraphics[width=8.4cm]{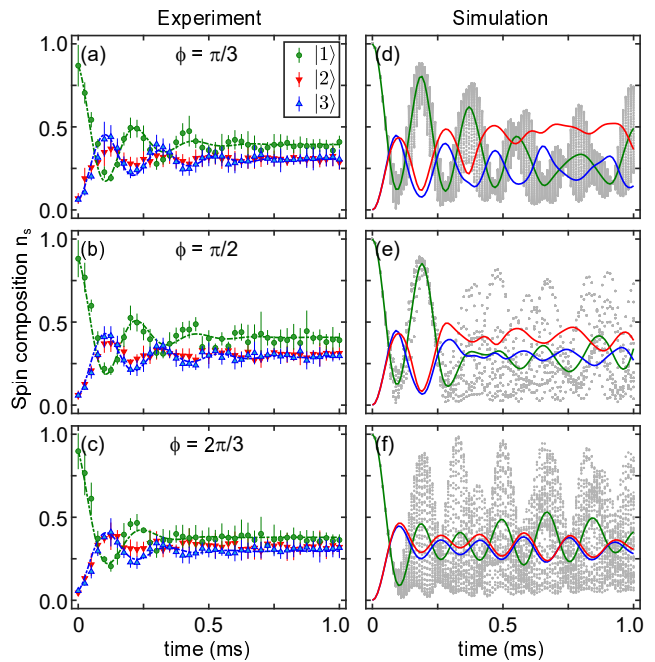}
\caption{Time dependence of the spin composition of periodic boundary synthetic Hall ladders for different values of effective magnetic flux (a) $\phi=\pi/3$, (b) $\pi/2$, and (c) $2\pi/3$. The dash-dotted line indicates fitted curve for corresponding spin compositions using Eq.~(\ref{eq:dampingfit}). (d)-(f) Numerical simulations accounting the effect of random flux $\Delta\varphi$ are shown. The solid lines represent the spin compositions averaged over a hundred 1D chains and the gray dots are evolutions of $n_1(t)$ for individual chains. Each data point comprises five independent measurements of the same experiment and the error bar is their standard deviation. \label{fig:periodic}}
\end{figure}
\end{center}

Finally, we repeat the quench experiment for the synthetic 3-leg Hall ladders under periodic boundary conditions. As seen in the trajectory change from the 2-leg to 3-leg ladder, the quench evolution in the periodic boundary Hall ladder will be much more complicated because of the circling motion of the atoms around the tube. In Fig.~\ref{fig:periodic}(a)-\ref{fig:periodic}(c), we display the time dependencies of $n_s(t)$ obtained for different magnetic fluxes. The oscillation is quickly damped with the time constant $t_d<300~\mu$s. To characterize the damping, we fit the data of the spin populations in $|1\rangle$ and $|3\rangle$ to a damped oscillator model,
\begin{equation}\label{eq:dampingfit}
n_s(t) = n_{s,0} + [n_s(0)-n_{s,0}] e^{-t/t_d}\cos(2\pi f t),
\end{equation}
where $n_{s,0}$ is the equilibrium value after a long time, and $f$ is the oscillation frequency. Table~\ref{tab:damping} summarizes the fitting results. In comparison with the open boundary Hall ladders, the periodic boundary systems feature a strong suppression of the oscillations with over a 6-times shorter damping time. In our experiment, the magnitude of the damping effect in the synthetic Hall ladder system is enhanced as $\phi$ increases.

The intense damping observed in the synthetic Hall ladders under periodic boundary condition can be attributed to fluctuations of the longitudinal magnetic flux $\phi_\mathrm{L}$ threading the Hall tube~\cite{Luo2020}. When an atom encircles the Hall tube at site $j$, i.e., a cyclic transition of $|j,1\rangle\rightarrow |j,2\rangle \rightarrow |j,3\rangle \rightarrow |j,1\rangle$ is made, a net phase of $\Phi_j= \Delta \vec{k}\cdot \vec{r}_j + \Delta \varphi$ with $\Delta\vec{k}=2\vec{k}_{r2}+\vec{k}_{r3}-3\vec{k}_{r1}=3k_R[(1-\cos \theta) \hat{x} -\sin \theta \hat{y}]$ and $\Delta \varphi=2\varphi_2+\varphi_3-3\varphi_1$ is acquired, where $\varphi_s$ is the relative phase of Raman laser beam $s$ and $\vec{r}_j$ is the position vector of site $j$. From the relation of $\Phi_j=\rho \phi j + \phi_\mathrm{L}$ described in Sec.~\ref{sec:model}, the relative phases of the Raman laser beams directly determine the threading magnetic flux, $\phi_\mathrm{L}=\Delta \varphi$. In our experiment, the relative phases between the Raman beams were not actively stabilized, and thus large fluctuations of $\Delta \varphi$ might lead to a damping effect in the experiment results, which involve averaging many independent measurements. Here, we remark that in the open boundary ladders, the relative phases between the Raman laser beams can be gauged away and have no active roles in the system dynamics. 

\begin{center}
\begin{table}[]
\centering
\caption{Measured $t_d$ (in ms) obtained from spin compositions $|1\rangle$ and $|3\rangle$ in synthetic Hall ladders for different boundary conditions and effective magnetic flux $\phi$. Numbers in brackets represent the 95\% confidence intervals from the fits.}
 \label{tab:damping}
\begin{tabular}{@{}c|c|c|c|c@{}}
\toprule\hline
                   		 &   $\phi$  & $\pi/3$ & $\pi/2$ & $2\pi/3$ \\ \midrule\hline
\multirow{2}{*}{Exp. $n_1$} & open 2-leg & 1692.1(265.2) & 1234.0(215.3) & 1480.2(309.3) \\ \cmidrule(lr){2-2}
                            & periodic & 141.3(30.6) & 145.3(33.3) & 102.5(22.0) \\ \hline \cmidrule(lr){2-2}
\multirow{2}{*}{Exp. $n_3$} & open 2-leg & 1873.1(280.7) & 1316.9(253.2) & 1461.4(305.3) \\ \cmidrule(lr){2-2}
                    		 & periodic & 226.6(50.0) & 210.7(46.1) & 145.2(37.7)  \\ \hline
\multirow{1}{*}{Sim. $n_1$} & periodic & 443.3(93.2) & 246.2(71.4) & 96.2(23.6) \\ \bottomrule\hline	
\end{tabular}
\end{table}
\end{center}

In Fig.~\ref{fig:periodic}(d)-\ref{fig:periodic}(f), we present the numerical simulation results for the quench dynamics with the periodic boundary conditions, including the random phase effect by averaging over 100 1D chains, while assuming evenly distributed $\phi_\mathrm{L}$. Our simulation explains well the damping behavior observed in the experiment. For a quantitative comparison with the experiment results, we determine the damping time by fitting the model curve of Eq.~(\ref{eq:dampingfit}) to the calculated $n_1$ at early times $t<0.5$~ms. The fitting results, listed in Table~\ref{tab:damping}, are found to be in reasonable agreement with the experiment data. As an example, a damping time of $t_d=96.2(23.6)$~ms for $\phi=2\pi/3$ is obtained, consistent with the experimentally measured value of $t_d=102.5(22.0)$~ms.

In Ref.~\cite{Luo2020}, Luo {\it et al.} pointed out that in our Raman beam setup, 1D tubes with different $y$ positions experience a different $\phi_\mathrm{L}$. Indeed, in the expression of $\Phi_j$, the wavevector $\Delta\vec{k}$ has a nonzero $y$-directional component, and thus the longitudinal magnetic flux $\phi_\mathrm{L}$ changes by $\Delta\vec{k}\cdot (d_y \hat{y})$ in the next tube for the $y$ direction. This means that the measurement in our experiment is intrinsically ensemble averaged even with $\Delta \varphi$ being perfectly controlled. In Fig.~\ref{fig:periodic}(d)-\ref{fig:periodic}(f), we plot the time evolution of $n_1$ of individual chains for the random distribution of $\phi_\textrm{L}$ as grey curves. Meanwhile, our experiment results shown in Fig.~\ref{fig:periodic}(a)-\ref{fig:periodic}(c) indicate that the variance of the measured value is much smaller than the expected value for a single tube experiment with a random $\phi_\mathrm{L}$, implying the averaging aspect of the system.

\section{Conclusion}\label{sec:conclusion}
We presented a synthetic Hall ladder system for atoms in a 1D optical lattice and demonstrated the control of the effective magnetic flux by investigating its quench dynamics for various ladder structures. For the open boundary ladders, we observed the characteristic spiral motions of the atoms, which were well explained by our numerical simulations based on the Bloch equations, including the effective damping. For the cylindrical Hall ladder, we observed a significantly strong damping in the quench evolution. We attributed this to fluctuations of the longitudinal magnetic flux $\phi_\mathrm{L}$ threading the Hall cylinder~\cite{Luo2020} and verified it through the numerical simulations of the experiment. The control of $\phi_\mathrm{L}$ could be improved in our setup by employing a 2D sample in the $xz$ plane or modifying the geometric configuration of the Raman beams. For example, we may rotate the irradiation directions of all Raman beams by $(\pi-\theta)/2$ in the $xy$ plane to make $\Delta \vec{k}$ aligned to the leg direction, resulting in a uniform $\phi_\mathrm{L}$ over the entire sample, which will be conducted as an immediate extension of this work. Combined with tunable atomic interactions based on the orbital Feshbach resonance~\cite{Pagano2015,Hofer2015}, we expect that the Hall ladder system provides interesting opportunities for investigating the correlation effects, such as the fractional charge behavior~\cite{Zeng2015}. Finally, we point out that when tunneling is allowed for a direction orthogonal to the leg, the 1D ladder system is transformed into a 2D multi-layer Hall system with an in-plane magnetic field~\cite{Hou2020}.

\section{Acknowledgments}
We thank Jin Hyoun Kang for the fruitful discussion. This work was supported by the National Research Foundation of Korea (2014-H1A8A1021987, NRF-2018R1A2B3003373, NRF-2019M3E4A1080399) and the Institute for Basic Science in Korea (IBS-R009-D1).


\begin{references}
\bibitem{OSG} For optical Stern-Gerlach detection, a laser pulse detuned by -860~MHz with respect to $|^1$S$_0,F=5/2\rangle$$\rightarrow$$|^3$P$_1,F'=7/2\rangle$ resonance is applied for 2~ms on the samples, followed by 8~ms of free expansion.
\bibitem{damping} The phenomenological damping is taken into account by $\langle k_e \rangle = (\langle k \rangle - \bar{\langle k \rangle})e^{-t/t_d} + \bar{\langle k \rangle}$, where $\bar{\langle k \rangle}$ is the mean value determined from the experiment and $t_d=1$~ms.
\end{references}
\end{document}